\documentclass{acmsiggraph}
\pdfoutput=1

\usepackage{xcolor}
\usepackage{bm}

\title{Stable and Controllable Neural Texture Synthesis and \\ Style Transfer Using Histogram Losses}

\author{Eric Risser$^1$, Pierre Wilmot$^1$, Connelly Barnes$^{1,2}$\\$^1$Artomatix, $^2$University of Virginia\\ \vspace{1ex}}
\pdfauthor{Eric Risser}

\TOGonlineid{0387}


\CopyrightYear{2017}
\setcopyright{acmcopyright}
\conferenceinfo{CONFERENCE PROGRAM NAME}{MONTH, DAY, and YEAR} 
\isbn{THIS-IS-A-SAMPLE-ISBN}\acmPrice{\$15.00}
\doi{http://doi.acm.org/THIS/IS/A/SAMPLE/DOI}

\newlength{\h}

\newcommand{\ignorethis} [1] {}

\newcommand{\sectnum    } [1] {\ref{#1}}
\newcommand{\sect       } [1] {Section~\sectnum{#1}}

\newcommand{\fignum     } [1] {\ref{#1}}
\newcommand{\fig        } [1] {Figure~\fignum{#1}}

\newcommand{\eqnnum     } [1] {\mbox{(\ref{#1})}}
\newcommand{\eqn        } [1] {equation~\eqnnum{#1}}

\definecolor{verydarkgreen}{rgb}{0,0.2,0}
\definecolor{verydarkorange}{rgb}{0.4,0.2,0}
\definecolor{verydarkyellow}{rgb}{0.55,0.55,0}

\ifdefined\ShowNotes
\newcommand{\Connelly}[1]{\colornote{verydarkgreen}{Connelly}{#1}}
\newcommand{\Eric}[1]{\colornote{dblue}{Eric}{#1}}
\newcommand{\Pierre}[1]{\colornote{maroon}{Pierre}{#1}}
\newcommand{\todo}[1]{\colornote{darkgreen}{Todo}{#1}}
\else
\newcommand{\Connelly}[1]{}
\newcommand{\Eric}[1]{}
\newcommand{\Pierre}[1]{}
\newcommand{\Yuting}[1]{}
\newcommand{\todo}[1]{}
\fi

\ifdefined\ShowNotes
  \newcommand{\colornote}[3]{{\color{#1}\bf{#2: #3}\normalfont}}
\else
  \newcommand{\colornote}[3]{}
\fi

\definecolor{darkred}{rgb}{0.7,0.1,0.1}
\definecolor{darkgreen}{rgb}{0.1,0.7,0.1}
\definecolor{verydarkgreen}{rgb}{0.0,0.5,0.0}
\definecolor{verydarkblue}{rgb}{0.0,0.0,0.7}
\definecolor{cyan}{rgb}{0.7,0.0,0.7}
\definecolor{dblue}{rgb}{0.2,0.2,0.8}
\definecolor{maroon}{rgb}{0.76,.13,.28}
\definecolor{burntorange}{rgb}{0.81,.33,0}

\definecolor{uhref_color}{HTML}{000066}

\newcommand\ulinec{\bgroup\markoverwith
      {\textcolor{uhref_color}{\rule[-0.5ex]{2pt}{0.4pt}}}\ULon}

\newcommand{\nsfurl}[1]{\ulinec{\small{\color{uhref_color}{\url{#1}}}}}

\newcommand{\styletransferfigs}{Figures \ref{fig:results_style_transfer_bonzai}-\ref{fig:results_style_transfer_shuttle}}

\begin{document}


 \teaser{
   \includegraphics[width=6.3in, trim=0.5in 2.25in 0.5in 2.6in]{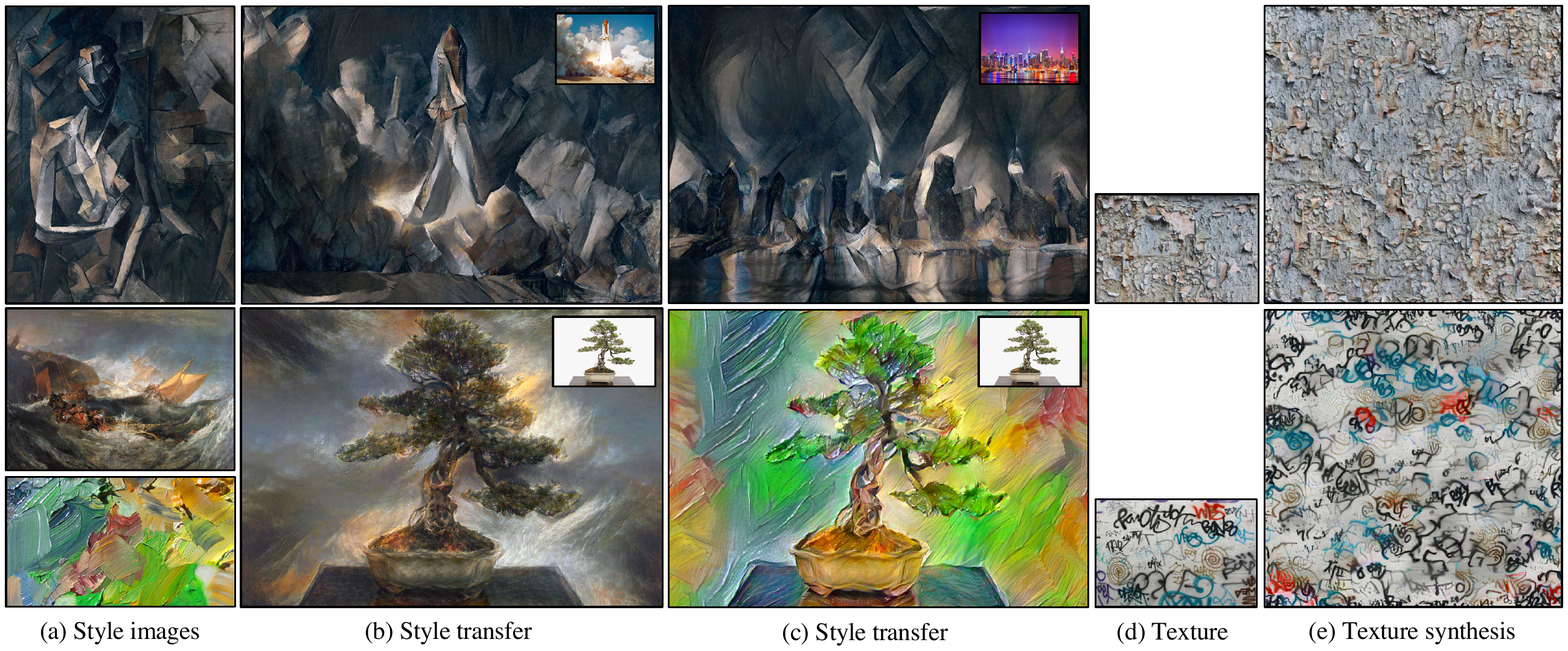}   
   \caption{Our style transfer and texture synthesis results. The input styles are shown in (a), and style transfer results are in (b, c). Note that the angular shapes of the Picasso painting are successfully transferred on the top row, and that the more subtle brush strokes are transferred on the bottom row. The original content images are inset in the upper right corner. Unless otherwise noted, our algorithm is always run with default parameters (we do not manually tune parameters). Input textures are shown in (d) and texture synthesis results are in (e). For the texture synthesis, note that the algorithm synthesizes creative new patterns and connectivities in the output.}
 }

\maketitle

\begin{abstract}

Recently, methods have been proposed that perform texture synthesis and style transfer by using convolutional neural networks (e.g. Gatys et al.~\shortcite{gatys2015texture,gatys2016image}). These methods are exciting because they can in some cases create results with state-of-the-art quality. However, in this paper, we show these methods also have limitations in texture quality, stability, requisite parameter tuning, and lack of user controls. This paper presents a multiscale synthesis pipeline based on convolutional neural networks that ameliorates these issues. We first give a mathematical explanation of the source of instabilities in many previous approaches. We then improve these instabilities by using histogram losses to synthesize textures that better statistically match the exemplar. We also show how to integrate localized style losses in our multiscale framework. These losses can improve the quality of large features, improve the separation of content and style, and offer artistic controls such as paint by numbers. We demonstrate that our approach offers improved quality, convergence in fewer iterations, and more stability over the optimization.


\end{abstract}

%
%


%
%


\keywordlist

\conceptlist

\printcopyright

\section{Introduction}

In recent years, deep convolutional neural networks have demonstrated dramatic improvements in performance for computer vision tasks such as object classification, detection, and segmentation~\cite{krizhevsky2012imagenet,he2016}. Because of the success of these models, there has also been much interest in adapting these architectures for synthesis tasks in graphics and computational photography. For instance, in computational photography, deep architectures have been used for many tasks, including editing of photos ~\cite{tsai2016sky,yan2016automatic,changautomatic}, objects~\cite{zhu2016generative}, image style transfer~\cite{gatys2016image}, texture synthesis~\cite{gatys2015texture}, new view synthesis~\cite{kalantari2016learning}, and image inpainting~\cite{pathakCVPR16context,yang2016high}.

In this paper, we specifically focus on the use of convolutional neural networks (CNNs) for style transfer and texture synthesis. Recently, Gatys~et~al.~\cite{gatys2015texture,gatys2016image} proposed parametric synthesis models for these problems, which utilize CNNs. For some inputs, particularly for style transfer, these models can result in quite successful, state-of-the-art results (see \styletransferfigs). However, we found as we investigated these methods in depth that they are subject to a number of limitations. These include limitations in stability, ghosting artifacts, the need for per-image parameter tuning, and challenges in reproducing large-scale features. Furthermore, these methods do not incorporate artistic controls such as painting by numbers~\cite{hertzmann2001image,ritter2006painting,lukavc2013painting,lukavc2015brushables}.

\begin{figure}
	\centering
	\setlength{\h}{1.3in}
	\setlength{\tabcolsep}{1pt}
	\begin{tabular}{ccc}
	\includegraphics[width=3.25in]{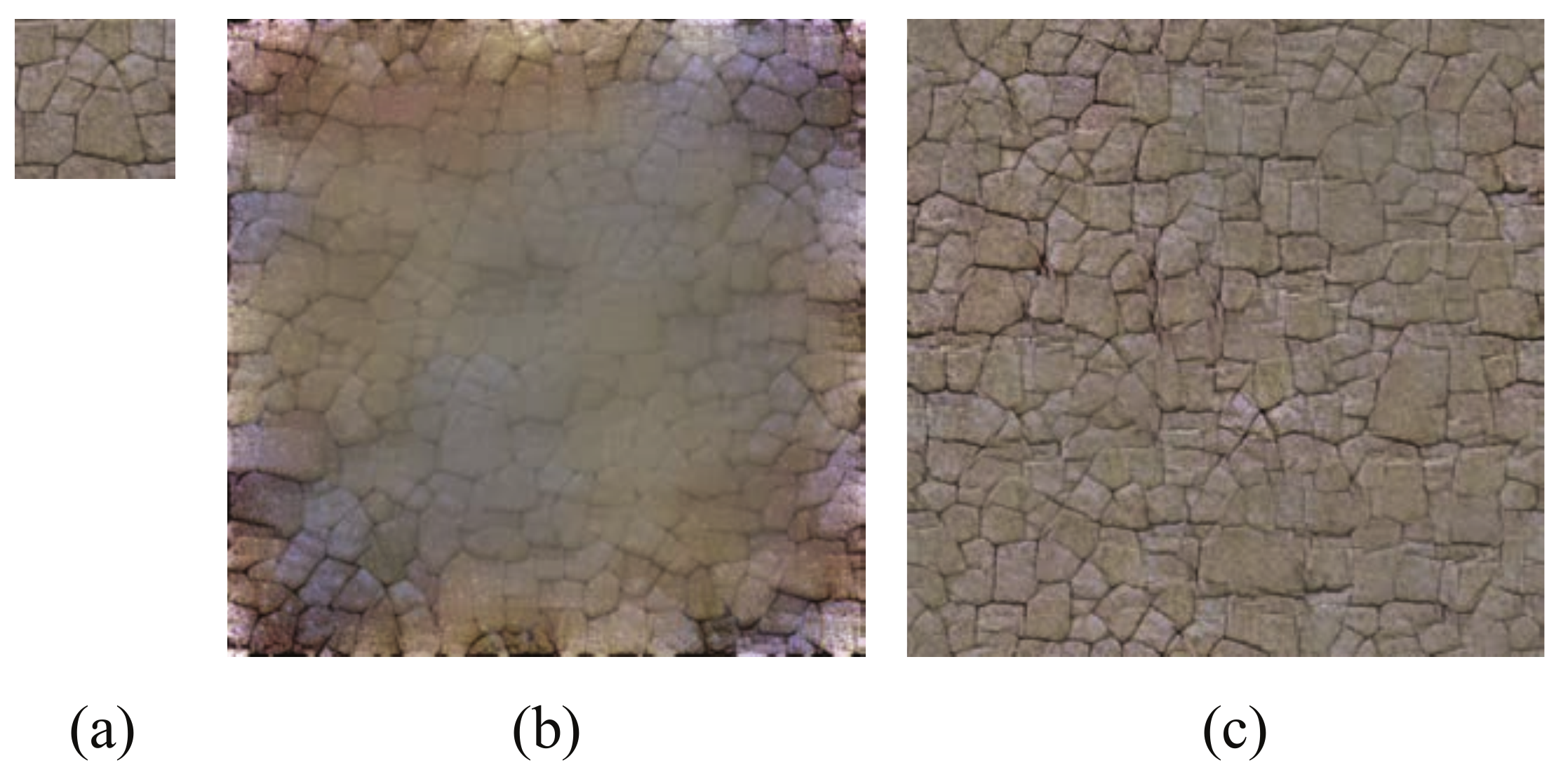}
	\end{tabular}
	\caption{Instabilities in texture synthesis of Gatys~et~al.~\protect\cite{gatys2015texture}. We show the input texture (a). A larger texture (b) is synthesized and shows a significant amount of instability where the brightness and contrast vary significantly throughout the image. By hand-tuning parameters, we can find parameter settings that produce pleasing results for this example (c). However, we still observe artifacts such as ghosting due to a smaller degree of instability.}
	\label{fig:unstable}
\end{figure}

Our first contribution in this paper is a demonstration of how and why such instabilities occur in the neural network synthesis approaches (see \sect{sec:problem}). Examples of instabilities are shown in \fig{fig:unstable}. In neural network synthesis methods such as Gatys~et~al~\shortcite{gatys2015texture,gatys2016image}, we have found that carefully tuning parameters on a per-image basis is often necessary to obtain good quality results, the optimization is often unstable over its iterations, and the synthesis process becomes more unstable as the size of the output increases.


We next demonstrate in \sect{sec:histogram} that such instabilities can be addressed by the use of novel histogram loss terms. These loss terms match not only the mean but also the statistical distribution of activations within CNN layers. We also show how to effectively minimize such histogram losses. Our histogram losses improve quality, reduce ghosting, and accelerate convergence. We initially formulate the histogram losses for texture synthesis and then in \sect{sec:style} discuss how to extend them for style transfer.

We next demonstrate in \sect{sec:localized} how quality and control can be further improved by using localized losses in a multiscale synthesis framework. In \sect{sec:localized_style}, we demonstrate how localized style losses can be used to incorporate artistic controls such as paint by numbers, as well as improve separation of content and style, and better reproduce large features. In \sect{sec:tuning}, we explain how we automatically select parameters, which removes the requirement for a human to manually tune parameters.

These contributions together allow us to achieve state-of-the-art quality for neural style transfer and for parametric neural texture synthesis.\footnote{Except for on regular textures, where Berger~et~al.~\shortcite{berger2016incorporating} have recently demonstrated a loss that improves regularity, which we do not currently include.} 


\section{Related work}

\textbf{Parametric texture synthesis.} Some early methods for texture synthesis explored parametric models. Heeger and Bergen~\shortcite{heeger1995pyramid} used histogram matching combined with Laplacian and steerable pyramids to synthesize textures. We are inspired by their use of histogram matching. Portilla and Simoncelli~\shortcite{portilla2000parametric} investigated the integration of many wavelet statistics over different locations, orientations, and scales into a sophisticated parametric texture synthesis method. These included cross-correlations between pairs of filter responses.

\textbf{Neural texture synthesis and style transfer.} In this paper, for short, we use ``neural" to refer to convolutional neural networks. Recently, Gatys~et~al.~\shortcite{gatys2015texture} showed that texture synthesis can be performed by using ImageNet-pretrained convolutional neural networks such as VGG~\cite{Simonyan14c}. Specifically, Gatys~et~al.~\shortcite{gatys2015texture} impose losses on co-occurrence statistics for pairs of features. These statistics are computed via Gram matrices, which measure inner products between all pairs of feature maps within the same layers of the CNN. The texture synthesis results of Gatys~et~al.~\shortcite{gatys2015texture} typically improve upon those of Portilla and Simoncelli~\shortcite{portilla2000parametric}. Gatys~et~al.~\shortcite{gatys2016image} later extended this approach to style transfer, by incorporating within CNN layers both a Frobenius norm ``content loss" to a content exemplar image, and a Gram matrix ``style loss" to a style exemplar image. We build upon this framework, and offer a brief review of how it works in the next section.  Concurrently to our research, Berger~et~al.~\shortcite{berger2016incorporating} observed that in the approach of Gatys~et~al., texture regularity may be lost during synthesis, and proposed a loss that improves regularity based on co-occurrence statistics between translations of the feature maps. Recently, Aittala~et~al.~\shortcite{aittala2016reflectance} used neural networks to extract SVBRDF material models from a single photo of a texture. Their method focuses on a more specific problem of recovering a SVBRDF model from a head-lit flash image. However, they do observe that certain instabilities such as non-stationary textures can easily result if sufficiently informative statistics are not used. We see this as connected with our observations about instabilities and how to repair them.  

\textbf{Feedforward neural texture synthesis.} Recently, a few papers~\cite{ulyanov2016texture,ulyanov2016instance,johnson2016}  have investigated the training of feedforward synthesis models, which can be pre-trained on a given exemplar texture or style, and then used to quickly synthesize a result using fixed network weights. The feed-forward strategy is faster at run-time and uses less memory. However, feed-forward methods must be trained specifically on a given style or texture, making the approach impractical for applications where such a pre-training would take too long (pre-training times of 2 to 4 hours are reported in these papers). 




\textbf{Non-parametric texture synthesis.} Non-parametric texture synthesis methods work by copying neighborhoods or patches from an exemplar texture to a synthesized image according to a local similarity term~\cite{efros1999texture,wei2000fast,lefebvre2005parallel,lefebvre2006appearance,kwatra2003graphcut,kwatra2005texture,barnes2009patchmatch}. This approach has also been used to transfer style~\cite{efros2001image,hertzmann2001image,barnes2015patchtable}. Some papers have recently combined parametric neural network models with non-parametric patch-based models~\cite{chen2016fast,li2016combining}. 

\section{A brief introduction to neural texture synthesis and style transfer}

In this section, we briefly introduce the texture synthesis and style transfer methods of Gatys~et~al.~\shortcite{gatys2015texture,gatys2016image}. For a more thorough introduction to these techniques, please refer to those papers. We show some example results from these methods later in the paper: neural texture synthesis is shown in \fig{fig:results_texsynth}, and style transfer results are shown in \styletransferfigs.

For texture synthesis~\cite{gatys2015texture}, we are given an input source texture $S$, and wish to synthesize an output texture $O$. We pass $S$ and $O$ through a CNN such as VGG~\cite{Simonyan14c}. This results in feature maps for the activations of the first $L$ convolutional layers, which we denote as $S_1 \ldots S_L$ and $O_1 \ldots O_L$. Then we minimize a loss $\mathcal{L}_\mathrm{gram}$ over the layers, which preserves some properties of the input texture by means of a Gram matrix:
\begin{equation}
\mathcal{L}_\mathrm{gram} = \sum_{l=1}^{L}  \frac{\alpha_l}{|S_l|^2} \|G(S_l) - G(O_l))\|_{F}^2
\label{eqn:gatys_tex}
\end{equation}
Here $\alpha_l$ are user parameters that weight terms in the loss, $|\cdot|$ is the number of elements in a tensor, $\|\cdot\|_F$ is the Frobenius norm, and the Gram matrix $G(F)$ is defined over any feature map $F$ as an $N_l \times N_l$ matrix of inner products between pairs of features:
\begin{equation}
G_{ij}(F) = \sum_{k}F_{ik}F_{jk}
\end{equation}
Here $F_{ij}$ refers to feature $i$'s pixel $j$ within the feature map. The synthesized output image $O$ is initialized with white noise and is then optimized by applying gradient descent to \eqn{eqn:gatys_tex}. Specifically, the gradient of \eqn{eqn:gatys_tex} with respect to the output image $O$ is computed via backpropagation.

Style transfer~\cite{gatys2016image} works similarly, but we are given a content image $C$, style image $S$, and would like to synthesize a stylized output  image $O$. We pass all three images through a CNN such as VGG, which gives activations for the first $L$ convolutional layers of $C_1 \ldots C_L$, $S_1 \ldots S_L$, $O_1 \ldots O_L$. Then the total style transfer loss combines the losses for the style image ($\mathcal{L}_\mathrm{gram}$) and the content image:
\begin{equation}
\mathcal{L}_\mathrm{transfer} = \mathcal{L}_\mathrm{gram} + \mathcal{L}_\mathrm{content}
\label{eqn:gatys_style}
\end{equation}
The content loss is a feature distance between content and output, which aims to make output and content look similar:
\begin{equation}
\mathcal{L}_\mathrm{content} = \sum_{l=1}^L\frac{\beta_l}{|C_l|} \|C_l - O_l\|_F^2
\label{eqn:gatys_content_loss}
\end{equation}
Again, $\beta_l$ are user weight parameters, and the output image $O$ is initialized with white noise and optimized using gradient descent.

\section{Our basic method using histogram losses}

In this section, we present our baseline texture synthesis and style transfer method, which incorporates histogram losses. We first briefly discuss in \sect{sec:useful_statistics} some statistics that can be useful for parameteric neural network synthesis. We then demonstrate in \sect{sec:problem} how and why instabilities occur in the previous approach of Gatys~et~al~\shortcite{gatys2015texture,gatys2016image}. We next explain in \sect{sec:histogram} how we address these instabilities using histogram losses, and how to effectively minimize histogram losses.

\subsection{Prelude: useful statistics for parametric synthesis}
\label{sec:useful_statistics}

\begin{figure}
	\centering
	\setlength{\h}{1.5in}
	\includegraphics[width=3.25in]{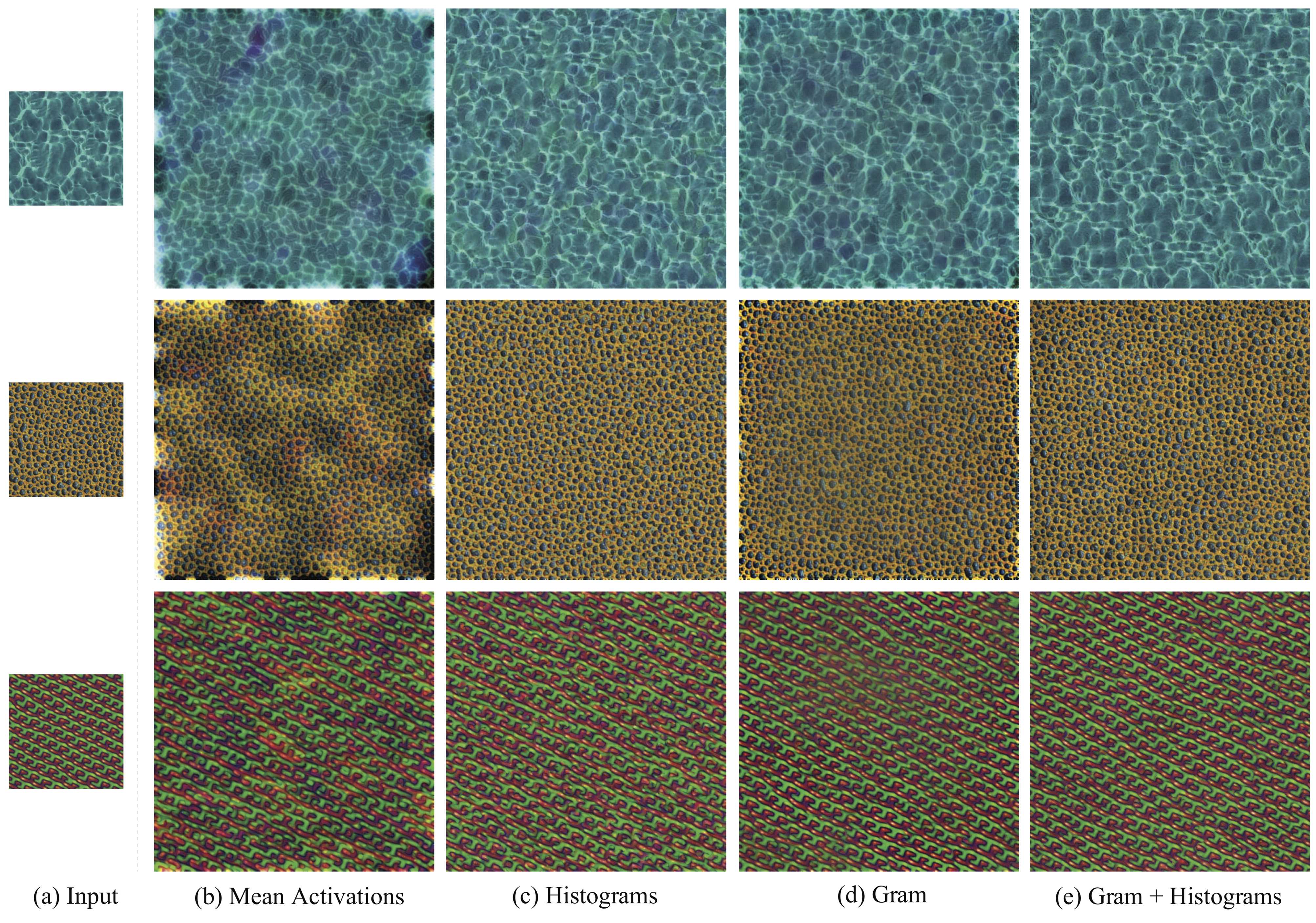}
	\caption{Different statistics that can be used for neural network texture synthesis. See the body of \sect{sec:useful_statistics} for discussion. \Eric{Replace this with two exemplars, four columns: mean activations, Gram matrices, histogram only, and using gram+histogram.}}
	\label{fig:gram_needed}
\end{figure}

If one revisits earlier parametric texture synthesis research such as Portilla and Simoncelli~\shortcite{portilla2000parametric}, one will quickly note that many different statistics can be used for texture synthesis. We investigated the effect of matching several such statistics for neural networks in \fig{fig:gram_needed}. We show the results of matching only the mean activations of each feature (with an $L^2$ loss), matching Gram matrices using \eqn{eqn:gatys_tex}, the full histogram matching that we develop later in \sect{sec:histogram}, and the combination of Gram matrices and our full histogram matching. Results with Gram matrix statistics tend to be generally better than results using mean activations. Results with histogram statistics are often more stable in terms of image intensity, but can break some small structures. The best results are obtained using our full method (Gram + histogram). Of course, additional statistics could also be investigated, such as the translational co-occurrence statistics of Berger~et~al.~\cite{berger2016incorporating}, but this is beyond the scope of our paper.

We find it interesting that although Gram matrices are widely used for synthesis~\cite{gatys2015texture,gatys2016image,selim2016painting,berger2016incorporating,johnson2016}, their results are often unstable. We investigate the causes of this instability in the next section. 

\subsection{The problem: instabilities}
\label{sec:problem}

\begin{figure}
	\centering
	\setlength{\h}{1in}
	\includegraphics[height=\h]{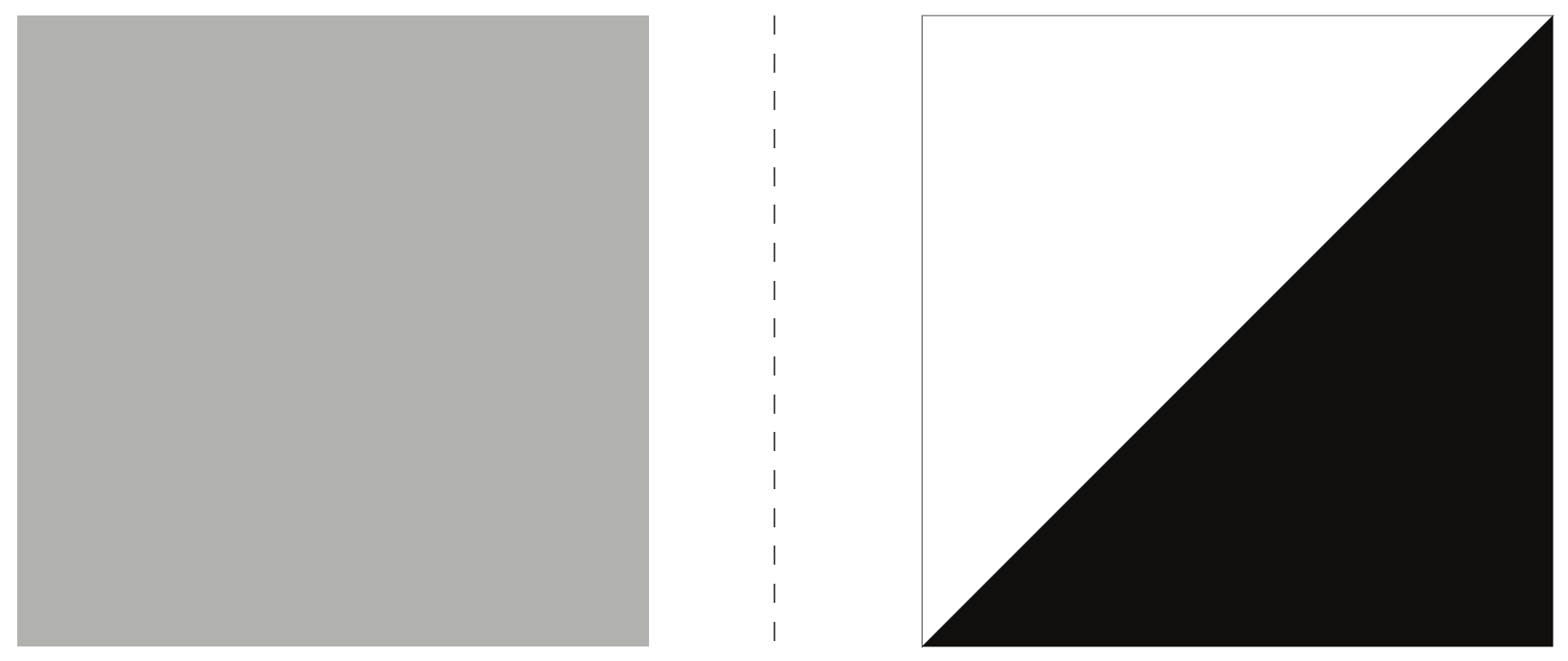} 
	\caption{At left, an example input image, which has a uniform distribution of intensities with a mean of $\mu_1 = 1/\sqrt{2} \approx 0.707$ and a standard deviation of $\sigma_1 = 0$. At right, an example output image, which has a non-uniform distribution with a mean of $\mu_2 = 1/2$ and a standard deviation of $\sigma_2 = 1/2$. If interpreted as the activation of a feature map with one feature, these two distributions have equivalent non-central second moments of $1/2$, and equal Gram matrices.}
	\label{fig:activation_problem}
\end{figure}

The instabilities are shown in \fig{fig:unstable}. The cause of these is quite simple to understand. To simplify and illustrate the problem, imagine that our input image is grey, as shown in \fig{fig:activation_problem} on the left. Our desired output from texture synthesis is thus also a grey image, with a mean value being $1/\sqrt{2}$ (about 70.7\% gray). The problem is that there are many distributions that will result in an equivalent Gram matrix, such as the output image on the right of \fig{fig:activation_problem}, where half of the image is 0 (black) and the other half is 1 (white). Of course, in reality, the Gram matrices do not match image intensities, but rather \emph{feature activations}, i.e. feature maps after applying the activation functions. However, the same argument applies: if we interpret the images of \fig{fig:activation_problem} as feature map activations, then activation maps with quite different means and variances can still have the same Gram matrix. 

It is slightly technical to understand this issue more formally. This is because the Gram matrix is statistically related to neither the mean nor covariance matrices, but instead to the matrix of non-central second moments. We now explain this formally. Let us consider the case of a feature activation map $F$ with $m$ features. In the rest of this section, for brevity, we will refer to ``feature map activations" simply as ``features," so all ``feature" always refers to the result of applying the activation function. We can summarize the statistics of the features in the feature map $F$ by using an $m$ dimensional random variable $\mathbf{X}$ to model the probability distribution of a given $m$-tuple of features. We can relate the random vector of features $\mathbf{X}$ and the feature map $F$. For example, if we normalize the Gram matrix $G(F)$ by the number of samples $n$, we obtain a sample estimator for the second non-central mixed moments $E[\mathbf{X}\mathbf{X}^T]$. Consequently, in the following discussion, we will sometimes informally refer to the (normalized) ``Gram matrix" and $E[\mathbf{X}\mathbf{X}^T]$ interchangeably, even though one is actually a sampled estimator for the other. For the following argument we thus set $\frac{1}{n}G(F) = E[\mathbf{X}\mathbf{X}^T]$. Define the mean feature $\bm{\mu} = E[\mathbf{X}]$. By a general property of covariance matrices, we have that $\Sigma(\mathbf{X}) = E[\mathbf{X}\mathbf{X}^T] - \bm{\mu}\bm{\mu}^T$, where $\Sigma$ indicates a covariance matrix. After rearranging, we obtain:
\begin{equation}
E[\mathbf{X}\mathbf{X}^T] = \Sigma(\mathbf{X}) + \bm{\mu}\bm{\mu}^T
\label{eqn:covar}
\end{equation}
For simplicity, let us now consider the case where we have only one feature, $m = 1$. By substituting into \eqn{eqn:covar}, we obtain:
\begin{equation}
\frac{1}{n}G(F) = E[X^2] = \sigma^2 + \mu^2 = \|(\sigma, \mu)\|^2
\label{eqn:covar1d}
\end{equation}
Here $\sigma$ is the standard deviation of our feature $X$. Suppose we have a feature map $F_1$ for the input source image, and a feature map $F_2$ for the synthesized output, and that these have respective feature distributions $X_1, X_2$, means $\mu_1, \mu_2$, and standard deviations $\sigma_1, \sigma_2$. Thus, from \eqn{eqn:covar1d}, we know that the  maps will have the same Gram matrix if this condition holds:
\begin{equation}
\|(\sigma_1, \mu_1)\| = \|(\sigma_2, \mu_2)\|
\label{eqn:covar_equality}
\end{equation}
We can easily use this to generate an infinite number of 1D feature maps with different variances but equal Gram matrix. Clearly this is bad for synthesis. Specifically, this means that even if we hold the Gram matrix constant, then \emph{the variance $\sigma_2^2$ of the synthesized texture map can be freely change} (with corresponding changes to the mean $\mu_2$ based on \eqn{eqn:covar_equality}), or conversely, that \emph{the mean $\mu_2$ of the synthesized texture map can freely change} (with corresponding changes to the variance, $\sigma_2^2$). This property leads to the instabilities shown in \fig{fig:unstable}. For simplicity, we assume that the CNN is flexible enough to generate any  distribution of output image features that we request. Suppose we wish to generate an output texture with a different variance (e.g. $\sigma_2 \gg \sigma_1$) but equal Gram matrix. Then we can simply solve \eqn{eqn:covar1d} for $\mu_2$, and obtain $\mu_2 = \sqrt{\sigma_1^2 + \mu_1^2 - \sigma_2^2}$. In fact, this is how we generated the different distributions with equal Gram matrices in \fig{fig:activation_problem}. The left distribution $X_1$ has $\mu_1 = 1/\sqrt{2}$ and $\sigma_1 = 0$, and we set a larger standard deviation $\sigma_2 = 1/2$ for the right distribution, so we obtain $\mu_2 = 1/2$.

In the multidimensional case $m > 1$, if there is no correlation between features, then we simply have $m$ separate cases of the previous 1D scenario. Thus, while maintaining the same Gram matrix, we can clearly change all of the variances however we like, as long as we make a corresponding change to the mean. This can clearly lead to instabilities in variance or mean.

However, in the multidimensional scenario, typically there are correlations between features. Consider the scenario where we are given an input feature map $F_1$, and thus the input feature random vector $\bm{X}_1$ and its mean $\bm{\mu}$ and covariance matrix $\Sigma(\bm{X}_1)$. Now, we wish to explore the set of output feature random vectors $\bm{X}_2$ with equal Gram matrix but different variance. In computer graphics, color adjustment models based on affine operations are often explored for finding correspondences~\cite{hacohen2011non}, and adjusting colors~\cite{siddiqui2008hierarchical}, including for texture synthesis~\cite{darabi2012image,diamanti2015synthesis}. Due to these affine models, we thought it appropriate to investigate whether synthesized textures will be stable if we hold their Gram matrix constant but otherwise allow features to undergo affine changes. We thus explored applying an affine transformation to our random vector of input feature activations $\bm{X}_1$ to obtain a transformed random vector of output feature activations $\bm{X}_2 = \bm{A}\bm{X}_1 + \bm{b}$, where $\bm{A}$ is an $m \times m$ matrix, and $\bm{b}$ is an $m$ vector. Then, using \eqn{eqn:covar}, we can set the Gram matrices of $\bm{X}_1$ and $\bm{X}_2$ equal to obtain:
\begin{equation}
\begin{split}
E[\bm{X}_2\bm{X}_2^T] & = \bm{A}\Sigma(\bm{X}_1)\bm{A}^T + (\bm{A}\bm{\mu}+\bm{b})(\bm{A}\bm{\mu}+\bm{b})^T  = \\
E[\bm{X}_1\bm{X}_1^T] & = \Sigma(\bm{X}_1) + \bm{\mu}\bm{\mu}^T.
\label{eqn:covar_expansion}
\end{split}
\end{equation}

We decided to constrain the variances of the output random feature activation vector $\bm{X}_2$ along the main diagonal of its covariance matrix, so that the variances are equal to a set of ``target" output image feature activation variances. We can then solve for the remaining unknown variables in the transformation matrix $\bm{A}$ and vector $\bm{b}$. Unfortunately, closed-form solutions of the resulting quadratic equations appear to generate multi-page long formulas, even for the two-dimensional case. However, intuitively, there are more unknowns than equations, so it should often be possible to generate a output feature distribution $\bm{X}_2$ with different variances than the input texture's feature distribution $\bm{X}_1$, but with the same Gram matrix. Specifically, there are $m(m+1)$ unknowns for $\bm{A}$ and $\bm{b}$, whereas there are only $m(m+3)/2$ constraints, due to \eqn{eqn:covar_expansion} ($m(m+1)/2$ constraints due to the upper half of the symmetric matrix, plus $m$ constraints for the known output feature variances).

To verify the intuition that it is possible to construct multidimensional distributions that are related by affine transformations and have different variance but equal Gram matrices, we ran extensive numerical experiments for solving the equations \ref{eqn:covar_expansion} in dimensions $m = 1 \ldots 16, 32, 64$. We found that in every case there was a solution. Specifically, we generated large numbers of mean and covariance matrices for the feature distributions of the input texture, and the output feature variances, with each sampled from the uniform distribution $U(0, 1)$.

We conclude that there are many ways to change the variance of an input texture without changing its Gram matrix. This leads to the instabilities shown in \fig{fig:unstable}.

We also note that recent papers on neural texture synthesis (e.g.~\cite{gatys2015texture,ulyanov2016texture,berger2016incorporating}) usually demonstrate outputs at the same resolution or slightly larger than the input. This is a departure from papers in computer graphics, which traditionally synthesized output texture at a significantly larger size than the input exemplar~\cite{efros1999texture,wei2000fast,lefebvre2005parallel,lefebvre2006appearance,kwatra2003graphcut,kwatra2005texture}. For smaller texture sizes, the instability artifact is more subtle and can be partially mitigated by manually tweaking learning rates and gradient sizes and which layers in the network contribute to the optimization process (but it can still be observed in results, e.g.~\cite{gatys2015texture,berger2016incorporating}). The instability artifact grows as the output texture is enlarged.



\subsection{Solution: Histogram Losses}
\label{sec:histogram}

As we have just discussed, previous neural texture synthesis models \cite{gatys2015texture,gatys2016image,berger2016incorporating} typically use Gram matrices to guide the synthesis process. This results in instabilities due to not providing guarantees that the mean or variance of the texture is preserved. One improvement could be to explicitly preserve statistical moments of various orders in the texture's activations. However, we took a further step and decided to just preserve the entire histogram of the feature activations. More specifically, we augment the synthesis loss with $m$ additional histogram losses, one for each feature in each feature map. These can be viewed as matching the marginal distributions of each feature's activations. We also incorporate a total variation loss~\cite{johnson2016}, which improves smoothness slightly in the output image.

Thus, our combined loss for texture synthesis is: 
\begin{equation}
\mathcal{L}_\mathrm{texture}^\mathrm{(ours)} = \mathcal{L}_\mathrm{gram} + \mathcal{L}_\mathrm{histogram} + \mathcal{L}_\mathrm{tv}
\label{eqn:tex_synth_ours}
\end{equation}
However, it is slightly subtle to develop a suitable histogram loss. Suppose we take the naive approach and directly place an $L^2$ loss between histograms of the input source texture $S$ and the output image $O$. Then this loss has zero gradient almost everywhere, so it does not contribute to the optimization process.

Instead, we propose a loss based on histogram matching. First, we transform the synthesized layerwise feature activations so that their histograms match the corresponding histograms of the input source texture $S$. This matching must be performed once for each histogram loss encountered during backpropagation. We then add a loss between the original activations and activations after histogram matching.

We simply use an ordinary histogram matching technique~\cite{wikipedia_histogram_matching} to remap the synthesized output activations to match the activations of the input source texture $S$. Let $O_{ij}$ be the output activations for convolutional layer $i$, feature $j$, and $O'_{ij}$ be the remapped activations. We compute the normalized histogram for the output activations $O_{ij}$ and match it to the normalized histogram for the activations of input source texture $S$, thus obtaining the remapped activations $O'_{ij}$. We repeat this for each feature in the feature map.




Once our output activations have been updated to mimic the input source texture, we find the Frobenius norm distance between the original activations and the remapped ones. Let $O_i$ be the activation map of feature map $i$ and $R(O_i)$ be the histogram remapped activation map. Then we have: 
\begin{equation}
\mathcal{L}_\mathrm{histogram} = \sum_{l=1}^L \gamma_l \| O_i - R(O_i) \|_F^2
\label{eqn:histo_loss}
\end{equation}
Here $\gamma_l$ is a user weight parameter that controls the strength of the loss.

To compute the gradient of this loss for backpropagation, we observe that the histogram remapping function $R(O_i)$ has zero gradient almost everywhere, and therefore can be effectively treated as a constant for the gradient operator. Therefore, the gradient of \eqn{eqn:histo_loss} can be computed simply by realizing $R(O_i)$ into a temporary array $O_i'$ and then computing the Frobenius norm loss between $O_i$ and $O_i'$.



\subsection{Extension to style transfer}
\label{sec:style}

The problem definition for style transfer can be thought of as a broadening of the texture synthesis problem. Texture synthesis is the problem of statistically resynthesizing an input texture. Style transfer is similar: one statistically resynthesizes an input style exemplar $S$ with the constraint that we also do not want the synthesized image $O$ to deviate too much from a content image $C$.

We have found that style transfer also suffers from the same instability artifacts we have shown in texture synthesis. Introducing histogram loss can offer the same benefits for this problem as well. Our style transfer strategy therefore follows Gatys~et~al.~\shortcite{gatys2016image}. We include both a per-pixel content loss and a histogram loss in the parametric texture synthesis equation. After including these, the overall style transfer loss becomes: 
\begin{equation}
\mathcal{L}_\mathrm{transfer}^\mathrm{(ours)} = \mathcal{L}_\mathrm{gram} + \mathcal{L}_\mathrm{histogram} + \mathcal{L}_\mathrm{content} + \mathcal{L}_\mathrm{tv}
\label{eqn:style_transfer_ours}
\end{equation}

\begin{figure}
	\centering
	\includegraphics[width=2.5in]{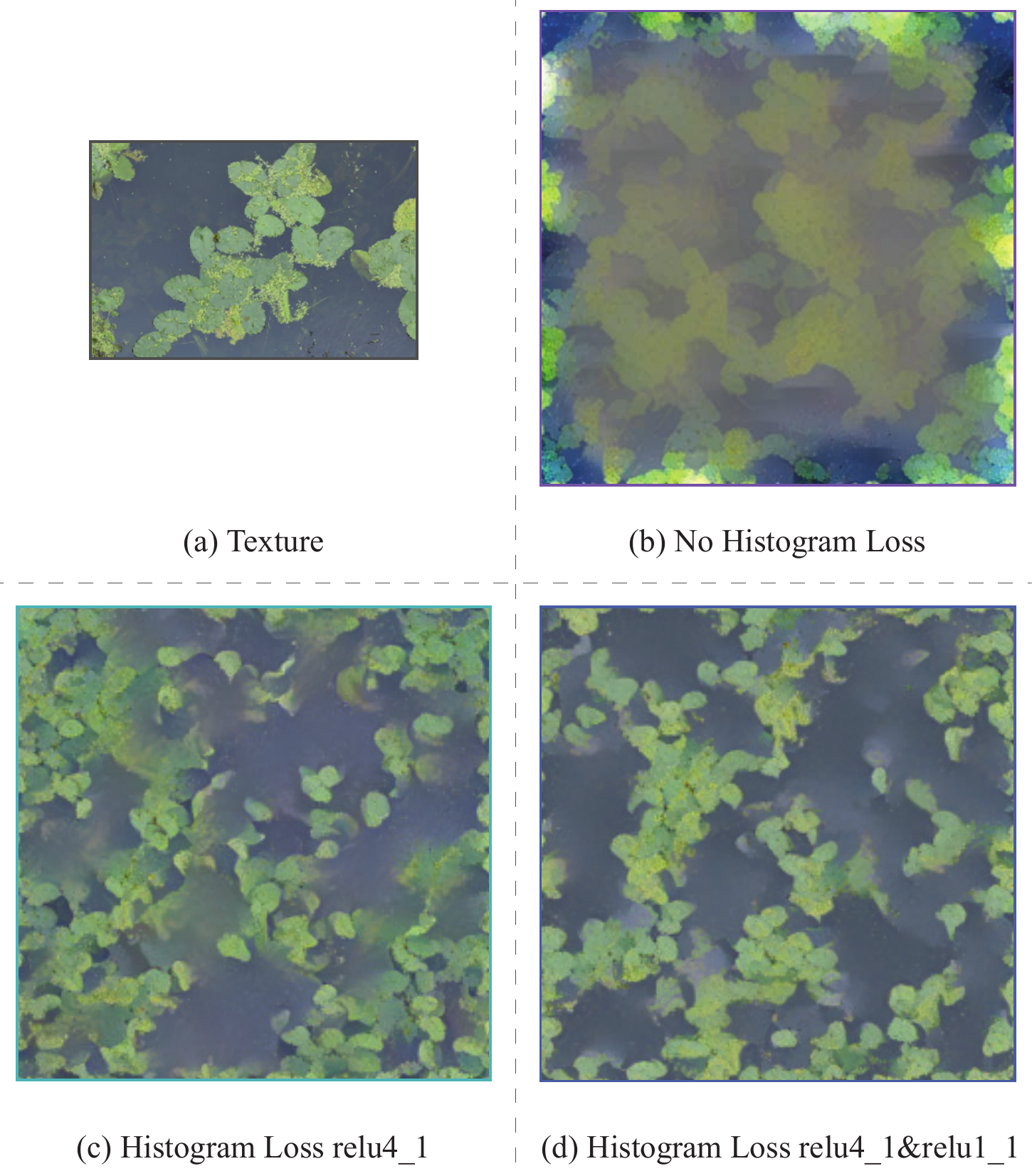}
	\setlength{\h}{1in}
	\caption{Instability and ghosting: in addition to instability problems, the baseline synthesis method~\protect\cite{gatys2015texture} using Gram matrices tends to interpolate sharp transitions in colors. As we discuss in \sect{sec:implementation}, our implementation uses the VGG-19 network~\protect\cite{Simonyan14c}. If we add a histogram loss (\eqn{eqn:histo_loss}) at later convolutional layers (rectified linear unit or ``relu" 4\_1 in VGG-19), this solves the instability issue, but the large overlapping receptive fields tend to blend features and create gradients where none exist in the input. If we also add a histogram loss to the first layer (relu 1\_1), this ameliorates this problem.\Connelly{write something about 1\_1 preventing ghosting.} \Eric{Need result here. Low priority figure.}}
	\label{fig:ghosting}
\end{figure}

\section{Localized losses for control and stability}
\label{sec:localized}

One concern with the parametric neural texture synthesis and style transfer approaches is that they do not include manual and automatic control maps that were previously used in non-parametric approaches~\cite{hertzmann2001image,barnes2009patchmatch,diamanti2015synthesis,fivser2016stylit}. We first introduce our multiresolution (pyramidal) synthesis approach. Then we introduce a localized style loss for adding artistic user controls.

\subsection{Multiresolution synthesis}
\label{sec:pyramid}

\begin{figure}
	\centering
	\includegraphics[width=3.25in]{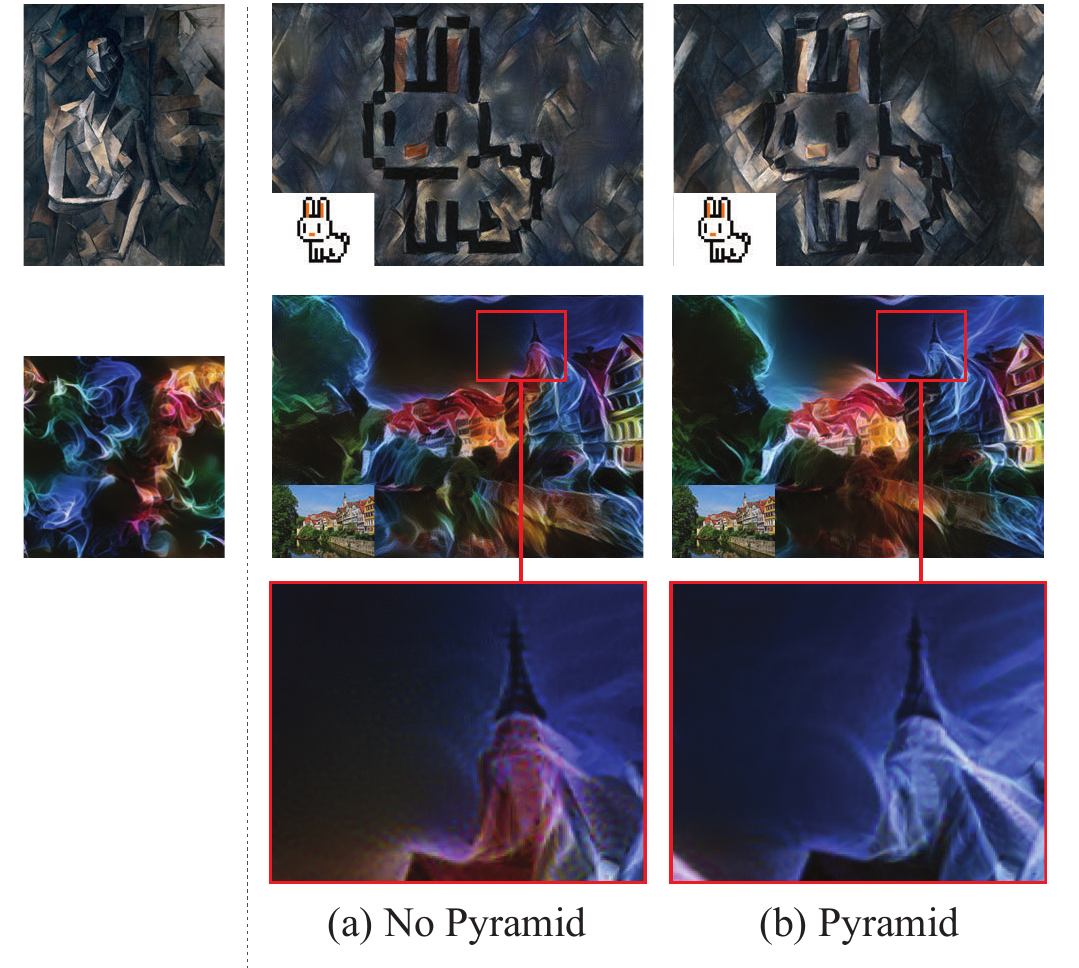}
	\setlength{\h}{1in}
	\caption{Comparison of pyramid and non-pyramid results. Style images are shown at left, and content images are shown inset. First row: pyramids blend coarse scale style features in with content features better. Second row: pyramids transfer coarse scale features better and reduce CNN noise artifacts. Third row: a zoom in from the second row, showing noise artifacts (at left) and better transfer of coarse-scale features (at right). \Eric{Need result here. Low priority figure.}}
	\label{fig:pyramid_no_pyramid}
\end{figure}

For both texture synthesis and style transfer, we have found results are generally improved by a coarse-to-fine synthesis using image pyramids~\cite{burt1983laplacian}. We use a ratio of two between successive image widths in the pyramid. We show in \fig{fig:pyramid_no_pyramid} a comparison of pyramid and non-pyramid results. We use pyramids for all results in this paper unless otherwise indicated. \Eric{Fix the preceding statements depending on what we actually select for our final algorithm.}

\subsection{Localized style loss for artistic controls}
\label{sec:localized_style}

Our approach for localizing loss functions is to make the observation that many images or “styles” are actually not a single texture but a collection of completely different textures laid out in a scene. As such it does not make sense to combine multiple textures into a single parametric model. Instead, we separate them out into multiple models. This approach has been known as ``painting by numbers" or ``painting by texture" in the literature~\cite{hertzmann2001image,ritter2006painting,lukavc2013painting,lukavc2015brushables}. This way of framing the problem has advantages for both texture synthesis and style transfer.

One way to do this would be to use multiple CNNs for different regions, and blend the regions together. However, that approach would be less efficient, and might introduce problems for the blending in transition regions. Instead, we perform painting by numbers synthesis in a single CNN.

\begin{figure*}
	\centering
	\includegraphics[width=5in]{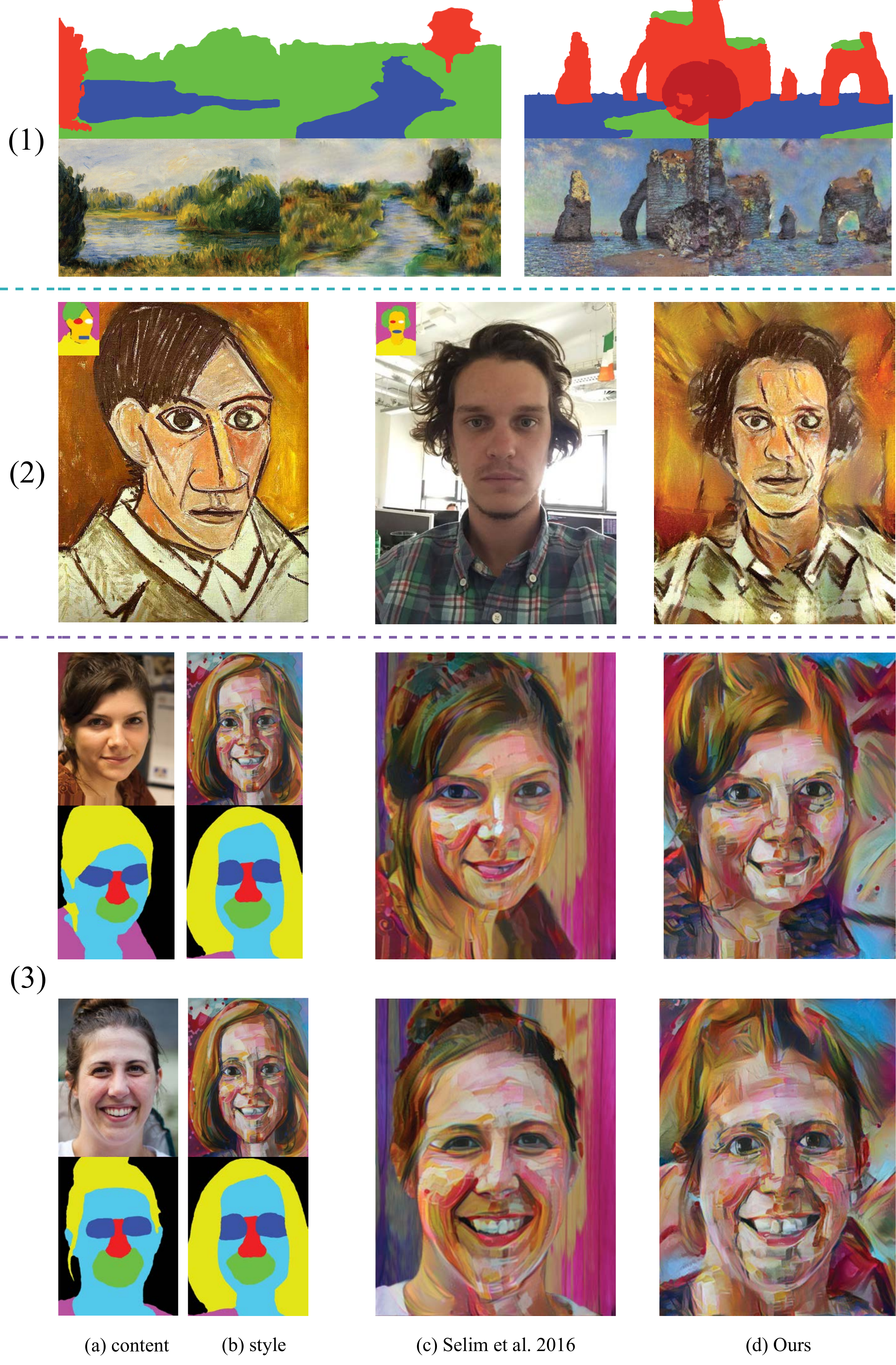}
	\caption{(1) Painting by Numbers: Here we show a first example of controllable parametric neural texture synthesis. Original images are on the left, synthesis results on the right, corresponding masks above each image. (2) One example of portrait style transfer using painting by numbers. (3) A second example of portrait style transfer. We show style transfer results for our method as compared to Selim~et~al.~\protect\shortcite{selim2016painting}. Shown are the content (a) and style image (b). For our method, these need to be augmented with indexed masks, as described in \sect{sec:localized_style}. In (c) and (d) are the results of Selim~et~al.~\protect\shortcite{selim2016painting} and our method. Note that our method preserves fine-scale artistic texture better. However, our method also transfers a bit more of the person's ``identity," primarily due to hair and eye color changes. Nevertheless, our method is not specialized for faces, so this is already an interesting result.}
	\label{fig:localized_masks}
\end{figure*}

As an input, the artist paints an ``indexed mask," where indices are painted on both the source texture (or style image) and the output image. \Eric{If we only do this for style transfer, rephrase the previous sentence, and add: We only investigated this effect for style transfer. Also rephrase the end of the first sentence in the next paragraph.} An example of these masks is shown in \fig{fig:localized_masks}. We assume there are $M$ indices.

Our localized style loss algorithm then proceeds by first collecting the pixels associated with each of the $M$ indexed regions within the texture or style exemplar $S$. For each region, we build a different Gram matrix and list of histograms.\footnote{One histogram is built for each feature, the same as in \sect{sec:histogram}.} We track the indices also on the synthesized output $O$, also tracking them as necessary through an image pyramid for coarse-to-fine synthesis. During synthesis, we modify our previous losses (Equations \ref{eqn:style_transfer_ours} and \ref{eqn:tex_synth_ours}) to be spatially varying. Specifically, we impose spatially varying Gram and histogram losses, where the style exemplar Gram matrices and exemplar histograms vary spatially based on the output index for the current pixel. For the histogram matching, we simply perform the matching separately within each of the $M$ regions defined by the indexed masks. Because the receptive fields for adjacent pixels overlap, backpropagation automatically performs blending between adjacent regions.

For style transfer it is important to note that often both the style and content images contain sets of textures that are semantically similar and should be transferred to each other. An example of this is shown in \fig{fig:localized_masks}. Using this approach we can transfer “higher level” style features such as eyes and lips, rather than just the “lower order” style features such as color and brush strokes. In \fig{fig:localized_masks} we also show a comparison of the results of our method against Selim~et~al.~\shortcite{selim2016painting}, which is a specialized method designed only for faces. For our method, we had to manually paint masks, unlike Selim~et~al.~\shortcite{selim2016painting}, however, our method can apply to many domains and is not specific to faces. 

\section{Automatic tuning of parameters}
\label{sec:tuning}

To obtain the best results, previous methods such as Gatys~et~al.~\shortcite{gatys2016image} would often manually tune parameters on a per-image basis. With our method, of course, the same tuning of parameters could be performed manually. However, this manual tuning process can be quite tedious. Thus, for all of our results, our method automatically determines the parameters. To ensure a fair comparison with previous approaches such as Gatys~et~al.~\shortcite{gatys2016image}, we also automatically determine their parameters in the figures throughout our paper, by selecting the default parameter values\footnote{We use Johnson's code~\shortcite{Johnson2015} for the comparisons with Gatys~et~al.~\shortcite{gatys2016image}}. We also show in a supplemental PDF a comparison of our automatic method with hand-tuned parameters for Gatys~et~al.~\shortcite{gatys2016image}. \Connelly{Make supplemental PDF.} We now describe our automatic parameter tuning process.

The parameters refer to the coefficients $\alpha_l$ in the Gram loss of \eqn{eqn:gatys_tex}, $\beta_l$ in the content loss of \eqn{eqn:gatys_content_loss}, $\gamma_l$ in the histogram loss of \eqn{eqn:histo_loss}, and a fourth parameter we call $\omega$ that is multiplied against the total variation loss~\cite{johnson2016}.

Our automatic tuning process is inspired by approaches such as batch normalization~\cite{ioffe2015batch}, which tune hyper-parameters during the training process so as to prevent extreme values for gradients. We thus dynamically adjust the parameters $\alpha_l$, $\beta_l$, $\gamma_l$, $\omega$ during the optimization process. Presently, our dynamic tuning is carried out with the aid of gradient information. We acknowledge that the optimization would likely be more mathematically well-founded if we instead dynamically tuned based on non-gradient information such as the magnitude of the losses or statistics within the losses. Nevertheless, this is what we currently do. During backpropagation, we encounter different loss terms $\mathcal{L}_i$, each of which has an associated parameter $c_i$ that needs to be determined ($c_i$ is one of the parameters $\alpha_l$, $\beta_l$, $\gamma_l$, $\omega$). We first calculate the backpropagated gradient $\mathbf{g}_i$ from the current loss term as if $c_i$ were 1. However, if the magnitude of $\mathbf{g}_i$ exceeds a constant magnitude threshold $T_i$, then we normalize the gradient $\mathbf{g}_i$ so its length is equal to $T_i$. We use magnitude thresholds of 1 for all parameters except for the coefficient $\alpha_l$ of the Gram loss, which has a magnitude threshold of 100. \Eric{Need Pierre to check the description in this paragraph, and fix if needed.}



\section{Implementation details}
\label{sec:implementation}

This section describes our implementation in more detail. Similarly to the previous literature, we also based our implementation on the VGG-19 network, which is pre-trained on the ImageNet dataset~\cite{Simonyan14c}. We use layers relu (rectified linear unit) 1\_1, relu 2\_1, relu 3\_1 and relu 4\_1 for the Gram losses in texture synthesis and style transfer. The histogram losses are computed only at layers relu 4\_1 and relu 1\_1 (other layers are ignored: this is equivalent to fixing their weights as zero in the loss). Content loss is computed only at relu 4\_1 when doing style transfer. The total variation loss smooths out noise that results from the optimization process and is thus performed only on the first convolutional layer.

As noted in \sect{sec:pyramid}, we synthesize our images in a multi-resolution process. We convert our input images into a pyramid. During synthesis, we start at the bottom of the pyramid, initialize to white noise, and after each level is finished synthesizing we use bi-linear interpolation to upsample to the next level.

Rather than adding a padding in our network, we instead opt for circular convolution. This produces textures that tile with themselves and does not seem to cause any strange effects during style transfer.


\section{Results and discussion}
\label{sec:results}

\begin{figure*}
	\centering
	\setlength{\h}{9in}
	\includegraphics[height=\h]{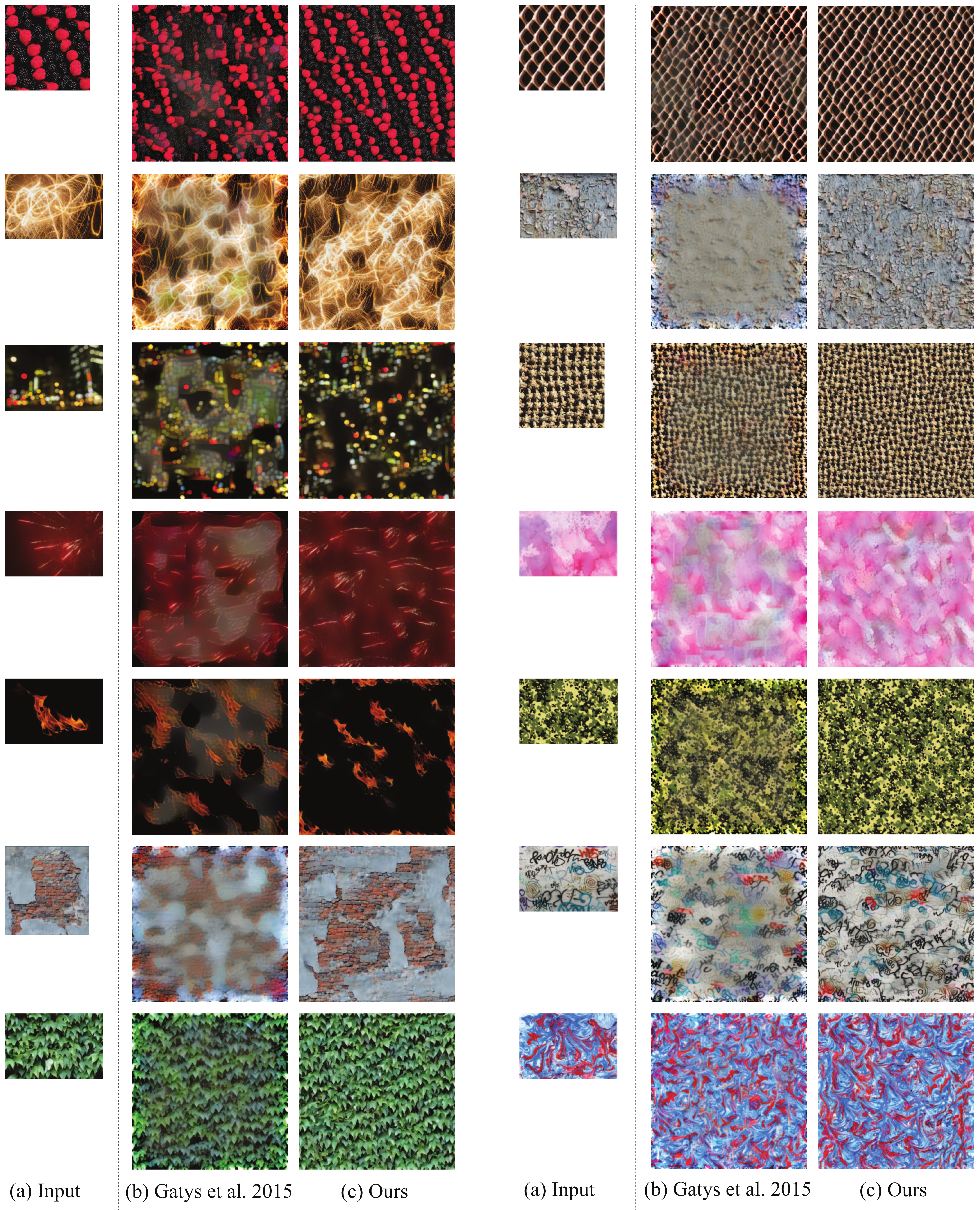}
	\caption{Results of our method for texture synthesis. Our Gatys results were generated from Johnson's code~\protect\cite{Johnson2015}. \Eric{This is a results placeholder! We love results! We should make more results!}}
	\label{fig:results_texsynth}
\end{figure*}

\begin{figure*}
	\centering
	\setlength{\h}{9in}
	\includegraphics[height=\h]{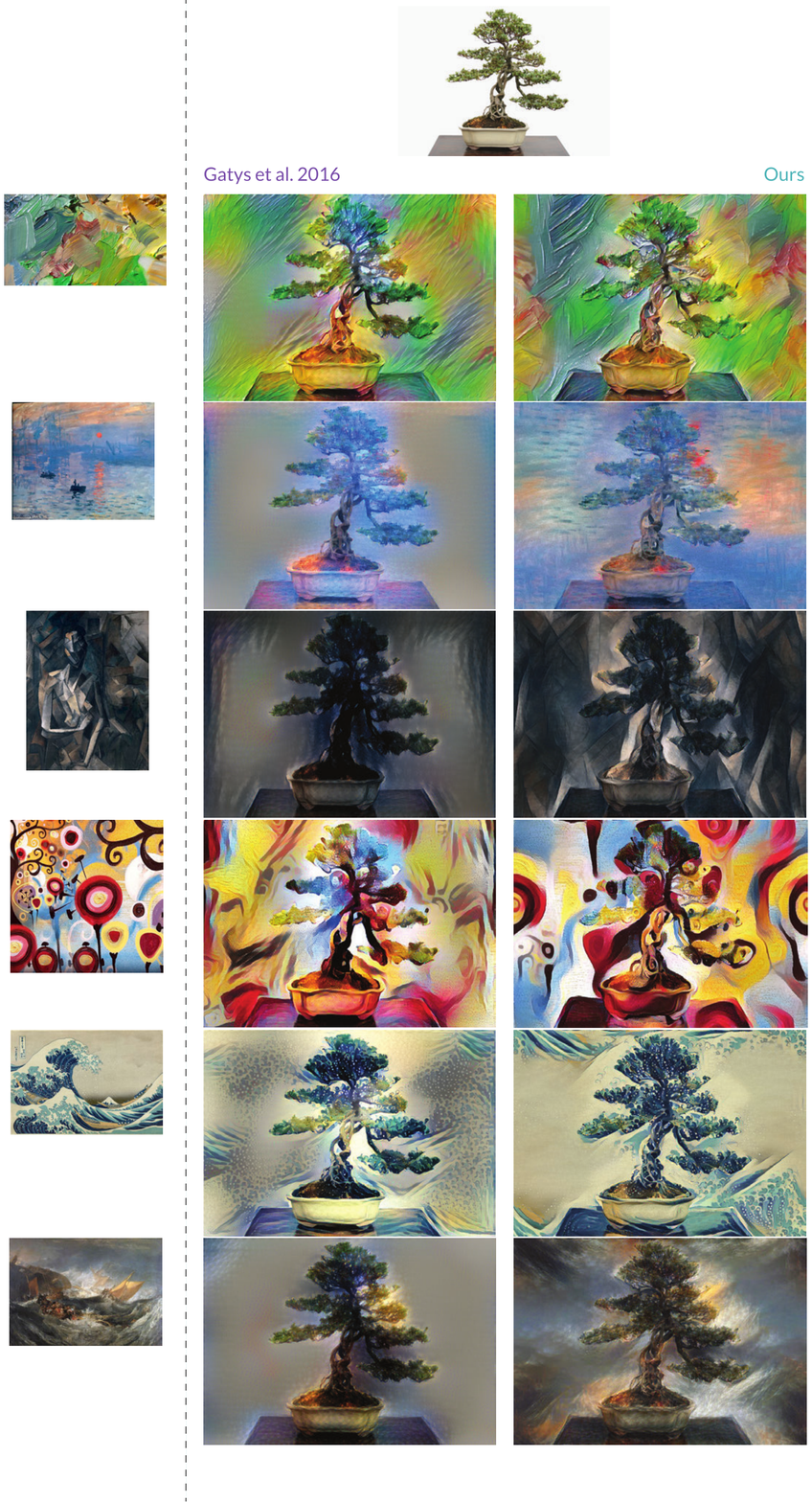}
	\caption{Results of our method for style transfer. Bonsai image from BonsaiExperience.com. \Eric{This is a results placeholder! We love results! We should make more results!}}
	\label{fig:results_style_transfer_bonzai}
\end{figure*}

\begin{figure*}
	\centering
	\setlength{\h}{9in}
	\includegraphics[height=\h]{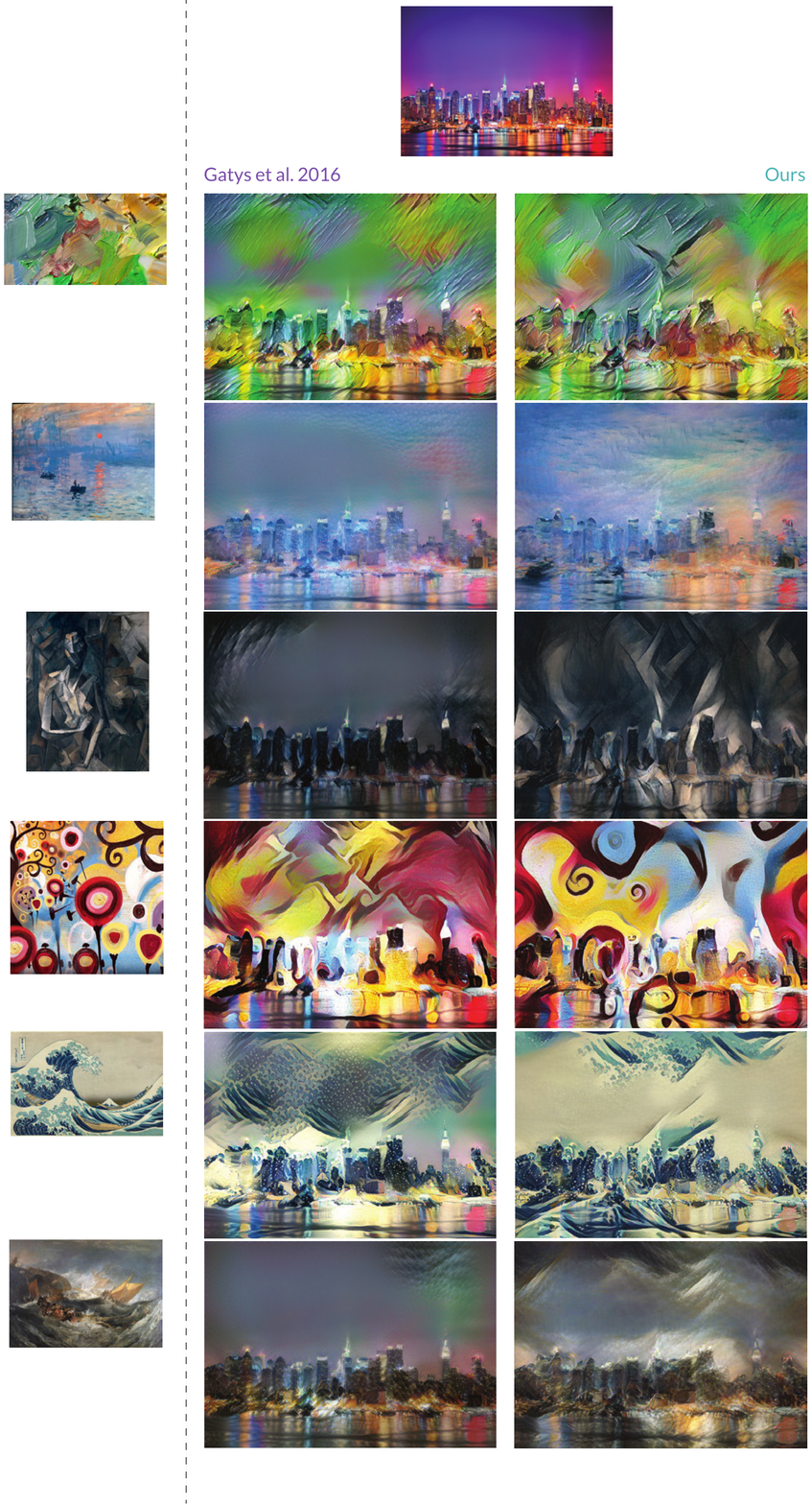}
	\caption{Results of our method for style transfer. Photo from Flickr user Wilerson S Andrade.\Eric{This is a results placeholder! We love results! We should make more results!}}
	\label{fig:results_style_transfer_NYC}
\end{figure*}

\begin{figure*}
	\centering
	\setlength{\h}{9in}
	\includegraphics[height=\h]{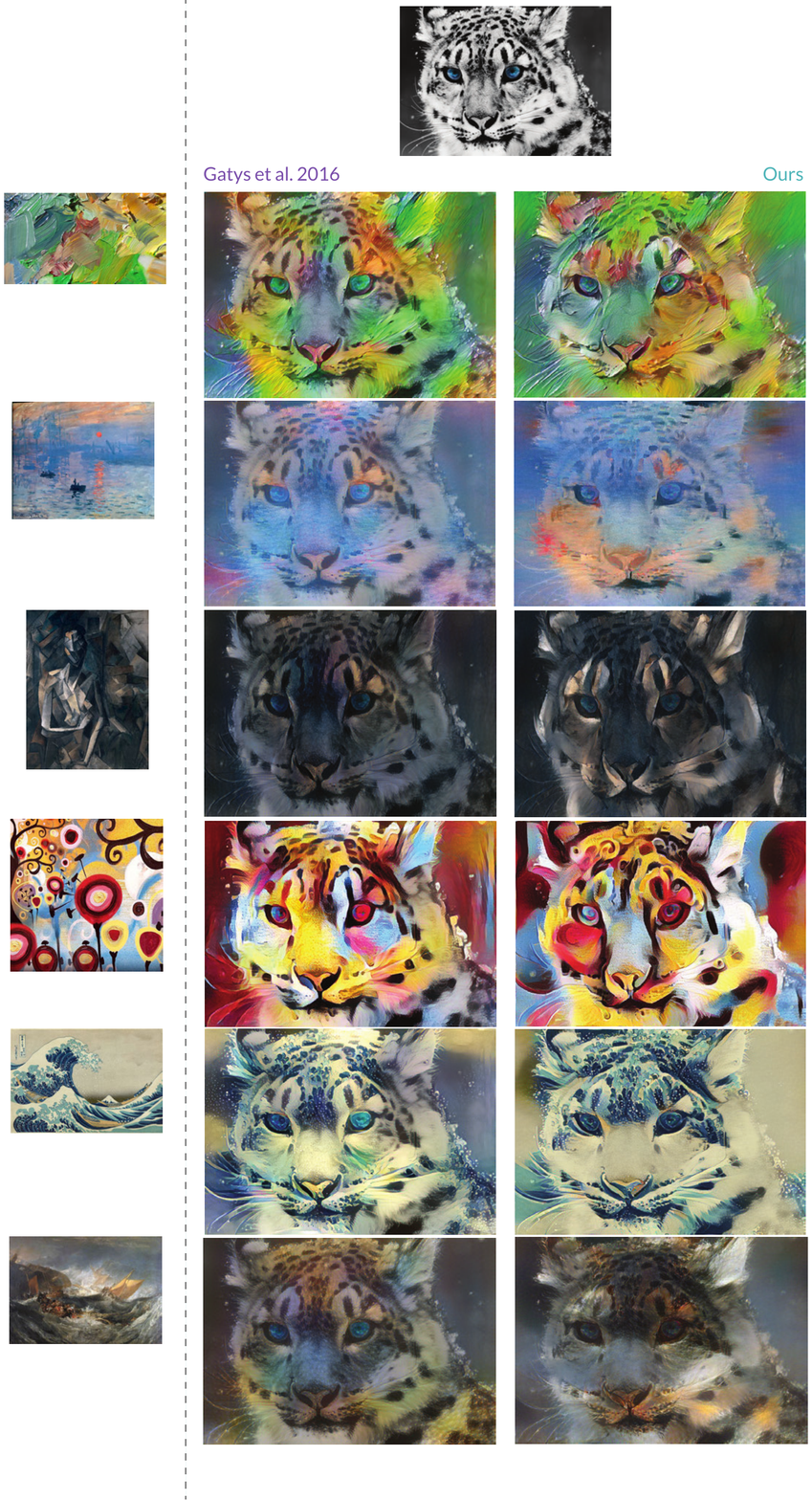}
	\caption{Results of our method for style transfer. \Eric{This is a results placeholder! We love results! We should make more results!}}
	\label{fig:results_style_transfer_tiger}
\end{figure*}

\begin{figure*}
	\centering
	\setlength{\h}{9in}
	\includegraphics[height=\h]{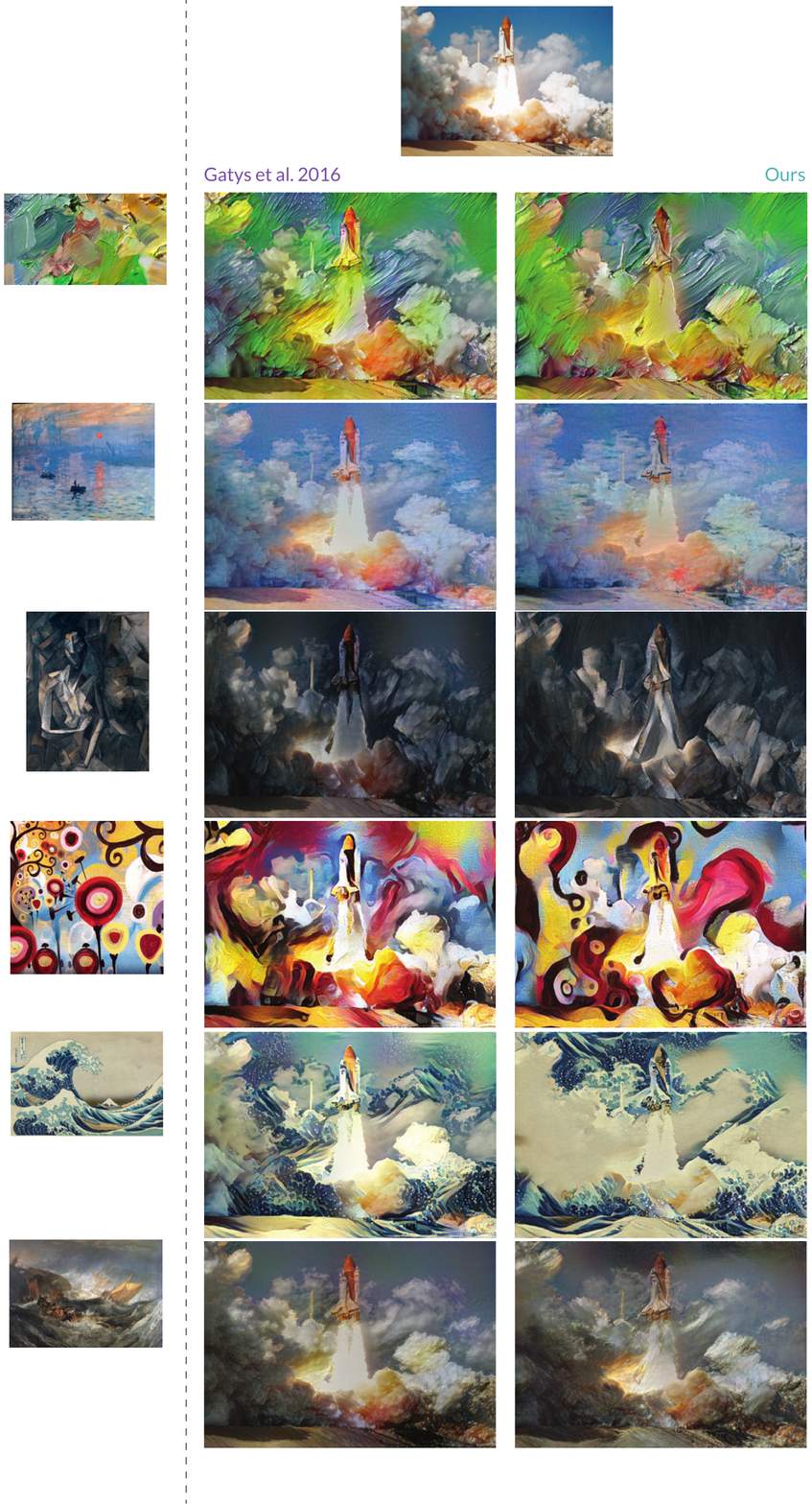}
	\caption{Results of our method for style transfer. Image from NASA.\Eric{This is a results placeholder! We love results! We should make more results!}}
	\label{fig:results_style_transfer_shuttle}
\end{figure*}

Results for our texture synthesis method are shown in \fig{fig:results_texsynth}. Our style transfer results are shown in \fig{fig:results_style_transfer_bonzai}, \fig{fig:results_style_transfer_NYC}, \fig{fig:results_style_transfer_tiger} and \fig{fig:results_style_transfer_shuttle}. Our supplemental HTML includes many more results.

We now discuss some advantages of our result. Our histogram loss addresses instabilities by ensuring that the full statistical distribution of the features is preserved. In addition to improving image quality, our full loss (including the histogram loss) also requires fewer iterations to converge. We use a mean of 700 iterations for our results, which we find consistently give good quality, and 1000 iterations for the results of Gatys~et~al~\shortcite{gatys2016image}, which we find is still sometimes unstable at that point. We note that methods based on Gram matrices such as Gatys~et~al~\cite{gatys2016image} can become unstable over the iteration count. We interpret this as being caused by the mean and variance being free to drift, as we discussed in \sect{sec:problem}. By adding histograms to our loss function, the result is more stable and converges better, both spatially and over iterations. 

Running times for our method are as follows. We used a machine with four physical cores (Intel Core i5-6600k), with 3.5 GHz, 64 GB of RAM, and an Nvidia Geforce GTX1070 GPU with 8 GB of GPU RAM, running ArchLinux. For a single iteration on the CPU, our method takes 7 minutes and 8 seconds, whereas Gatys~et~al.~\shortcite{gatys2016image} takes 15 minutes 35 seconds. This equates to our method requiring only 45.7\% of the running time for the original Gatys method. Our approach used three pyramid levels and histogram loss at relu4\_1 and relu1\_1. These metrics were measured over 50 iterations synthesizing a 512x512 output. We currently have most but not all of our algorithm implemented on the GPU. Because not all of it is implemented on the GPU, a speed comparison with our all-GPU implementation of Gatys~et~al.~\shortcite{gatys2016image} is not meaningful, therefore we ran both approaches using CPU only.



\section{Conclusion}

The key insight of this paper is that the loss function introduced by Gatys et al.~\shortcite{gatys2015texture} and carried forward by follow-up papers can be improved in stability and quality by imposing histogram losses, which better constrain the dispersion of the texture statistics. We also show improvements by automating the parameter tuning, and in artistic controls. This paper improves on these aspects for both texture synthesis and style transfer applications. 




\bibliographystyle{acmsiggraph}
\nocite{*}
\bibliography{paper}
\end{document}